# Design of Skyrmion Bags with Tunable Topology in Symmetry-Broken 2D Lattices

Junhuang Yang, Kaiying Dou, Ying Dai[*], Baibiao Huang, and Yandong Ma[*]

School of Physics, State Key Laboratory of Crystal Materials, Shandong University, Shandanan Str. 27, Jinan 250100, China

E-mail: daiy60@sdu.edu.cn (Y.D.); yandong.ma@sdu.edu.cn (Y.M.)

**Abstract**

Magnetic skyrmion bags, as high-order topological swirling spin textures, offer rich fundamental physics and distinct advantages for spintronic applications; however, their realization remains a formidable challenge, especially in two-dimensional (2D) systems. Here, through model analysis, we propose a novel design principle for engineering skyrmion bags with tunable topology in symmetry-broken 2D ferromagnetic lattices. The physics correlates to the delicate interplay of isotropic exchange interaction, Dzyaloshinskii-Moriya interaction and magnetic anisotropy, which can stabilize a rich variety of high-order topological spin states as well as the intriguing skyrmionium with zero topological charge. We further validate this mechanism in monolayer $CrInTe_2$ using first-principles calculations and atomistic spin model simulations, revealing the existence of field-free skyrmion bags. Furthermore, we find that a weak magnetic field triggers a transition to skyrmioniums that exhibit remarkable thermal stability up to 240 K. Our results provide a compelling platform for exploring high-order topological magnetism.

**Keywords:** skyrmion bags, skyrmionium, Dzyaloshinskii-Moriya interaction, magnetic anisotropy, first-principles.

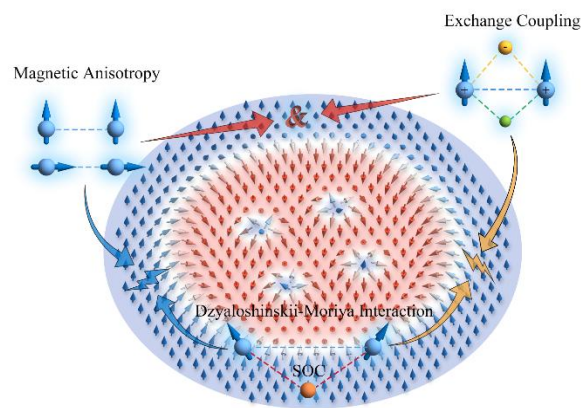

**Introduction**

Topological magnetism manifesting as chiral quasiparticles with nontrivial topology, represents a frontier in condensed matter physics due to its dual significance in fundamental science and spintronic applications [1-4]. These topologically protected configurations offer distinct advantages, including high stability and energy-efficient manipulability [5-7], which are essential for future high-density storage and computing technologies [8-10]. Beyond their device potential, such chiral spin textures serve as a rich playground for novel physics, enabling the exploration of exotic transport phenomena like the topological Hall effect, topological spin Hall effect, topological layer Hall effect and skyrmion Hall effect (SkHE) [11-13].

The topology of these chiral quasiparticles is quantified by the topological charge $Q$. While current research in this field has primarily focused on elementary skyrmion with $|Q| = 1$ [14,15], high-order swirling spin textures with $|Q| \neq 1$, such as skyrmion bags, are attracting increasing attention [16-18]. Typically, a skyrmion bag consists of a large outer skyrmion enclosing multiple inner skyrmions [19,20]. Beyond preserving the intrinsic properties of conventional skyrmions, skyrmion bags offer distinct advantages for spintronics, including significantly larger topological Hall effects [21] and more significant anisotropic magnetoresistance [22]. Meanwhile, the complex internal structures of such high-order textures promise to unveil richer fundamental physics [23]. Despite their potential, skyrmion bags remain largely unexplored, especially in two-dimensional (2D) systems. Crucially, a general design principle for engineering these states is still lacking, which not only limits our understanding of high-order topological magnetism but also hinders the systematic exploration of suitable material platforms.

In this Letter, we address this challenge by establishing a theoretical framework that enables the construction of skyrmion bags with tunable topology within symmetry-broken 2D ferromagnetic (FM) lattices. Our model analysis elucidates that the physics of these multiconfigurational states relies critically on the specific competition of isotropic exchange interaction, Dzyaloshinskii-Moriya interaction (DMI) and magnetic anisotropy. To demonstrate the feasibility of this scheme, we perform first-principles and atomistic simulations on monolayer $CrInTe_2$, which confirm the emergence of field-free high-order topologies. Crucially, our investigation reveals a field-driven evolution from skyrmion bags to topologically neutral skyrmioniums, a phase that maintains structural integrity up to 240 K. These findings provide the missing link between theoretical design and the material realization of high-order topological magnetism.

## Results and Discussion

Physically, the stabilization of topological magnetism in 2D lattices arises from the delicate energetic competition between the Heisenberg exchange interaction, the DMI and the magnetic anisotropy. While the antisymmetric DMI prefers a chiral twisting of magnetic moments to lower the energy, thereby destabilizing the collinear order, the isotropic exchange interaction and magnetic anisotropy impose an energy penalty on spatial variations, favoring rigid alignment along a specific quantization axis. In the regime where DMI dominates, the system proliferates chiral domain, condensing into a maze-like spin spiral (SS) ground state. Conversely, when the stiffening forces of exchange and anisotropy prevail, the continuous rotation of spins is suppressed, typically leading to a FM state or, under specific conditions, compact metastable excitations such as isolated skyrmions.

In principle, with bridging this gap, the evolution from a dense spin spiral state to the isolated skyrmion regime involves the topological breaking of labyrinthine domains. It is within this intermediate transition region that isolated skyrmions can nucleate and become encapsulated within larger domain boundaries, giving rise to high-order topologies known as skyrmion bags [**Fig 1**(a)]. Unlike elementary skyrmions, the formation of skyrmion bags requires a precarious balance: the domain walls must be flexible enough to loop into multi-soliton configurations yet rigid enough to prevent collapse. Consequently, these states typically emerge only within a vanishingly narrow window of the magnetic phase diagram, explaining their scarcity in both theoretical predictions and experimental observations [24,25]. In the following, we demonstrate that the synergistic enhancement of exchange interaction, DMI and magnetic anisotropy can effectively expand this stability window, rendering the deterministic design of skyrmion bags experimentally accessible.

To quantitatively verify this design principle and explore the stability of the resultant textures, we employ the following 2D atomistic spin Hamiltonian model:

$$H = -J \sum_{<i,j>} \mathbf{S}_i \cdot \mathbf{S}_j - K \sum_i (\mathbf{S}_i^z)^2 - \sum_{<i,j>} \mathbf{D}_{ij} \cdot (\mathbf{S}_i \times \mathbf{S}_j) \tag{1}$$

Here, $\mathbf{S}_i$ denotes the normalized spin vector localized at the $i^{th}$ magnetic atom. $<i, j>$ indicates summation over nearest-neighbor (NN) pairs. The parameters $J$, $K$ and $\mathbf{D}_{ij}$ represents the isotropy Heisenberg exchange, single-ion anisotropy and DMI, respectively. Topologically, a skyrmion bag acts as a composite soliton, comprising a primary outer skyrmion boundary that encapsulates $N$ inner skyrmions of opposite polarity. This configuration results in a net topological charge $Q = N - 1$. In the continuum limit, this charge is a topological invariant defined as [26]:

$$Q = \frac{1}{4\pi} \int \mathbf{m} \cdot \left( \frac{\partial \mathbf{m}}{\partial x} \times \frac{\partial \mathbf{m}}{\partial y} \right) dxdy \qquad (2)$$

where **m** denotes the normalized vector of spin. While the topological charge $Q$ provides a rigorous classification, it is structurally more intuitive to classify these bags by their internal multiplicity. Therefore, we adopt the notation $S(N)$—where $N = Q + 1$—to directly specify the total number of nested inner skyrmions. For instance, as illustrated in **Fig 1**(a), a skyrmion bag carrying a charge of $Q = 3$ encapsulates four inner skyrmions and is explicitly labeled as $S(4)$.

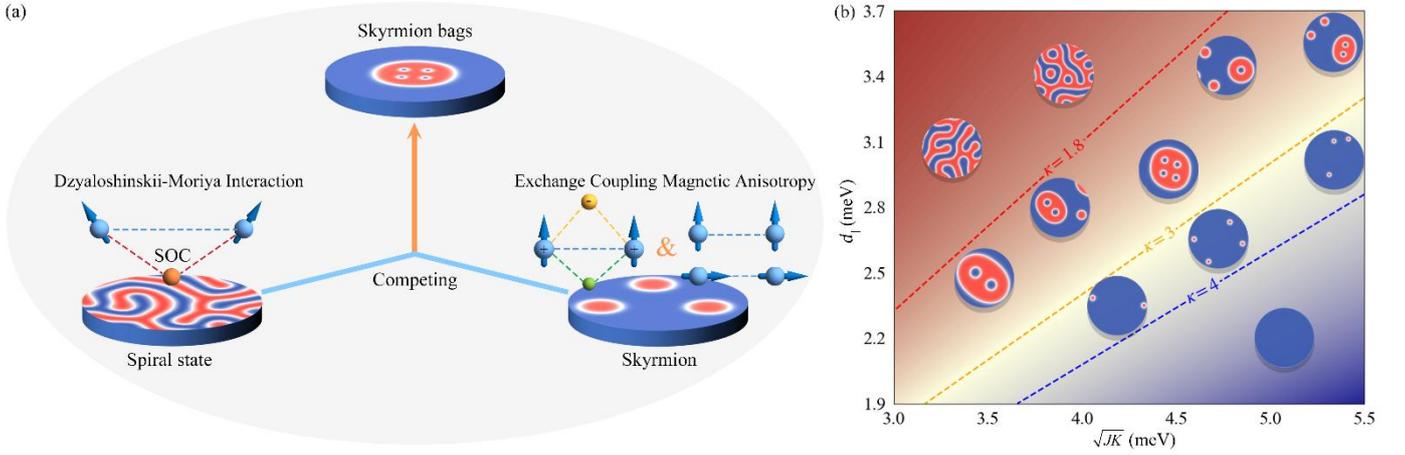

**Fig 1** (a) Schematic diagrams of the formation mechanism of skyrmion bags with $S(4)$. (b) Rescaled phase diagram described by the dimensionless parameter $\kappa$ [**Eq. (3)**]. The dashed lines indicate the critical boundaries for the spin spiral ($\kappa < 1.8$), skyrmion bags ($1.8 < \kappa < 3$), isolated skyrmions ($3 < \kappa < 4$) and the FM state ($\kappa > 4$). The insets highlight representative spin textures stabilized with each regime.

**Fig S1** presents the comprehensive phase diagram of the ground-state spin textures as a function of the interaction parameters $d_\parallel$ and $\sqrt{JK}$. The diagram clearly demarcates distinct regimes governed by the competition between chiral and collinear interactions. In the upper-left region, where the $d_\parallel$ dominates over the $\sqrt{JK}$, the system minimizes its energy by stabilizing a dense, maze-like spin spiral state. Conversely, in the lower-right region characterized by large $\sqrt{JK}$ and moderate $d_\parallel$, the penalty for spin rotation prevails, enforcing an FM order.

The intriguing physics emerges at the boundary between these two phases. As the $\sqrt{JK}$ is slightly reduced from the FM limit, isolated skyrmions begin to nucleate. More importantly, in the intermediate transition region where $d_\parallel$ and $\sqrt{JK}$ are comparable, the fierce competition leads to the topological breaking of labyrinth domains, giving rise to skyrmion bags. The stability of these high-order states exhibits a clear trend along the diagram's main diagonal. Specifically, as the system evolves towards the upper-left (decreasing $\sqrt{JK}$ and increasing $d_\parallel$), the outer domain wall of the bag expands, accommodating a varying number of inner skyrmions. Consequently, the topological charge $Q$ decrease, eventually reaching $Q = 0$—a state known as skyrmionium—before the system melts into the spiral phase.

The critical insight from **Fig S1** is the "synergistic enhancement" effect: simultaneously increasing $d_\parallel$ and $\sqrt{JK}$ significantly can broaden the stability window of skyrmion bags. This finding confirms our design principle, suggesting that stronger interactions render the deterministic creation of these textures experimentally accessible. Moreover, such wide stability window also facilitates the modulation of the topology of skyrmion bags through external magnetic field.

To provide a universal criterion for this phase behavior, we characterize the interplay of $d_\parallel$ and $\sqrt{JK}$ using a dimensionless parameter $\kappa$, which can be expressed as [27-29]:

$$\kappa = (\frac{4}{\pi})^2 \frac{2JK}{3d_\parallel^2} \tag{3}$$

**Fig 1**(b) remaps the phase diagram in terms of $\kappa$. The evolution of spin textures following a distinct hierarchy: the spin spiral state persists for $\kappa < 1.8$. In the window of $1.8 < \kappa < 3$, the spin spiral evolves into magnetic skyrmions, simultaneously giving rise to the long-sought skyrmion bags. When $\kappa$ exceeds 3, the multi-soliton bags vanish, leaving only elementary isolated skyrmions. Finally, for $\kappa > 4$, these skyrmions also annihilate and the system transitions into a trivial FM state. This dimensionless analysis reinforces that maximizing the absolute interaction strengths while maintains an optimal ratio $\kappa$ is the key for the material realization of the proposed skyrmion bags with tunable topology. This design principle provides clear guidance for selecting and engineering suitable material systems to achieve skyrmion bags with tunable topology.

Guided by this criterion, we identity monolayer CrInTe$_2$ as one candidate material for realizing skyrmion bags with tunable topology. **Fig 2**(a) presents the crystal structure of monolayer CrInTe$_2$. It crystallizes in a triangular lattice belonging to the space group $P3m1$. The unit cell comprises a Cr layer sandwiched between two Te layers, capped by an In layer in a Te-Cr-Te-In stacking sequence. Crucially, this specific atomic arrangement breaks the structural inversion symmetry, which is the fundamental prerequisite for generating the DMI required for chiral magnetism. Our calculations yield an optimized lattice constant of a = 4.03 Å, in excellent agreement with the previous work [30]. Furthermore, the dynamic and thermal stability of the monolayer is robustly verified by the absence of imaginary modes in phonon spectra and by ab initio molecular dynamics (AIMD) simulations (**Fig S2**).

To elucidate the microscopic origin of the magnetism, we analyze the electronic configuration of the Cr ions. The valence electronic configuration of Cr atom is $3d^54s^1$. In monolayer CrInTe$_2$, each Cr atom transfers three electrons to the surrounding Te atoms during bonding, leading to a final configuration of $3d^34s^0$. In the distorted octahedral coordination environment, the $d$ orbitals split roughly into two groups, i.e., a higher-energy doublet $e_g$ and a lower-energy triplet $t_{2g}$ orbitals. The three left electrons of the Cr atom will half-fill the $t_{2g}$ orbital, which would result in a magnetic moment of 3 $\mu_B$, as illustrated in **Fig S3**. Consistent with this picture, our spin polarized calculations show that the total magnetic moment of monolayer CrInTe$_2$ is 3.11 $\mu_B$ per unit cell, which is localized primarily on Cr atoms. Furthermore, the electronic band structure [**Fig 2(d)**], demonstrates that the Fermi level intersects only the spin-up bands, while the spin-down channel remains insulating with a band gap. This specific band topology confirms the half-metallic behavior of monolayer CrInTe$_2$.

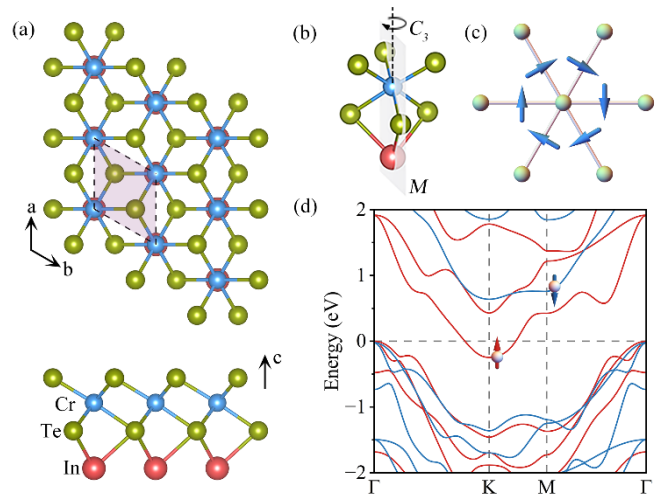

**Fig 2** (a) Crystal structure of monolayer CrInTe$_2$ from top and side views. (b) Distorted octahedral geometry of Cr atom. $M$ and $C_3$ represents the $M$ and $C_3$ symmetry, respectively. (c) DMI vectors between the nearest-neighboring Cr atoms (d) Band structure of monolayer CrInTe$_2$ without SOC.

To parameterize the effective spin Hamiltonian [**Eq. (1)**], we first extract the Heisenberg exchange interaction and magnetic anisotropy strengths by mapping the total energies of four distinct magnetic configurations (see **Fig S4**) to the model. The calculated NN isotropy Heisenberg exchange parameter $J$ is 34.12 meV, indicating a robust FM exchange coupling between adjacent Cr atoms. This behavior is consistent with the Goodenough−Kanamori−Anderson (GKA) rules [31,32], which predict FM super-exchange for the nearly 90° bond geometry inherent to this edge-sharing octahedral network. Furthermore, the single-ion anisotropy is determined to be $K = 0.86$ meV. The positive sign of $K$ confirms that the easy magnetization axis lies along the out-of-plane (OP) direction, favoring perpendicular magnetic anisotropy.

We then address the DMI in monolayer CrInTe$_2$, the pivotal ingredient for stabling topological magnetism. Symmetry analysis based on the $C_{3v}$ point group of monolayer CrInTe$_2$ [**Fig 2**(b)] reveals that the DMI is constrained by the three vertical mirror symmetry operations (*M*). According to Moriya's rules [33], the DMI vector between NN Cr atoms can be decomposed as $\mathbf{D}_{ij} = d_{\parallel}(\mathbf{u}_{ij} \times \mathbf{z}) + d_{\perp}\mathbf{z}$, where $\mathbf{u}_{ij}$ and **z** are the unit vectors from sites *i* to *j* and along the **z** direction, respectively. Due to the $C_{3v}$ symmetry, the OP DMI component $d_{\perp}$ cancel on average; consequently, the chirality is governed exclusively by the in-plane (IP) component $d_{\parallel}$.

To quantify this interaction $d_{\parallel}$, we consider the clockwise (CW) and anticlockwise (ACW) spin-spiral configurations (**Fig S5**). Intriguingly, a substantial intrinsic DMI of $d_{\parallel} = 3.94$ meV is obtained for monolayer CrInTe$_2$. The positive sign signifies a preferred CW chirality, as shown in **Fig 2**(c). To elucidate the microscopic origin of such a large DMI in monolayer CrInTe$_2$, we analyze the atomistic resolved spin-orbit coupling (SOC) energy difference $\Delta E_{SOC}$. As shown in **Fig S6**, the heavy Te atoms dominate the DMI contribution, consistent with the Fert-Levy mechanism [14,34]. Crucially, the two inequivalent Te sites exhibit opposing contributions: the Te-1 (the atom farther from the In atom) provide a dominant positive contribution that overrides that negative contribution from Te-2, thereby establishing the robust net DMI.

The derived magnetic parameters provide a striking validation of our design strategy based on synergistic enhancement. In monolayer CrInTe$_2$, both the effective anisotropy scale $\sqrt{JK}$ value (5.42 meV) and the DMI strength (3.94 meV) substantially surpass those of established topological magnets, such as MnSeTe [14] ($\sqrt{JK} = 2.21$ meV, DMI = 2.14 meV) and Co$_2$NF$_2$ [35] ($\sqrt{JK} = 1.71$ meV, DMI = $-1.01$ meV). Most critically, combining these values yields a dimensionless stability parameter of $\kappa = 2.04$ (see **Table S1**). This value falls squarely within the predicted window ($1.8 < \kappa < 3$) for stable skyrmion bags identified in our phase diagram. Consequently, monolayer CrInTe$_2$ is expected for hosting stable skyrmion bags with tunable topology.

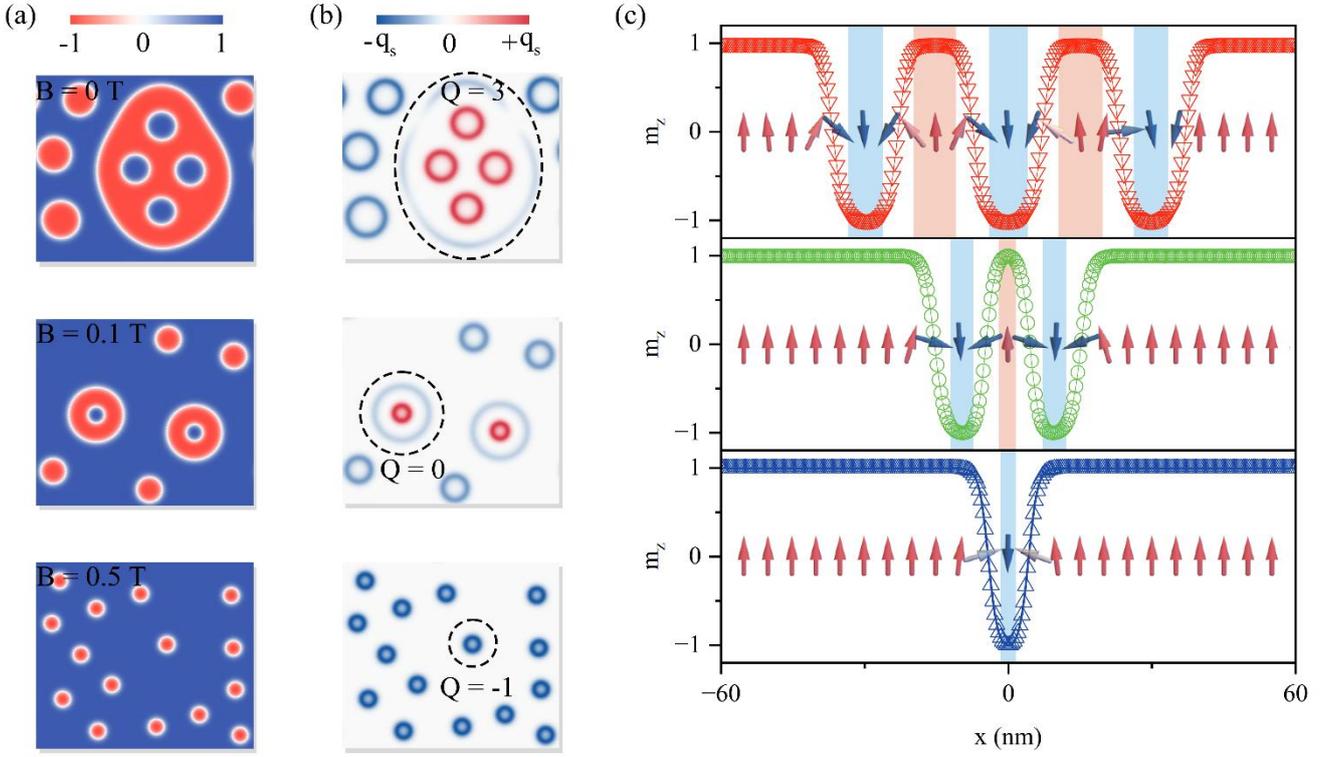

**Fig 3** Field-driven topological evolution in monolayer CrInTe$_2$. (a) Spin textures, (b) corresponding topological charge density maps and (c) cross-sectional profiles of the OP spin component passing through the geometric center of the structures. The columns correspond to the skyrmion bag $S(4)$ at zero field (B = 0 T), the skyrmionium at B = 0.1 T, and the elementary skyrmion at B = 0.5 T, respectively.

Utilizing these magnetic parameters, we investigate the spin textures in monolayer CrInTe$_2$ by conducting atomistic spin model simulations. **Fig 3**(a) illustrates the spin textures of monolayer CrInTe$_2$. Intriguingly, a spontaneous skyrmion bag is observed at 0 K without applying external magnetic field. The skyrmion bag comprises a large outer skyrmion boundary and four inner skyrmions with opposite polarity. Topological analysis in **Fig 3**(b) reveals a net topological charge $Q = 3$. This integer value arises from the superposition of the outer boundary's charge $Q = -1$ and the contribution from four inner skyrmions with $Q = +1$, consistent with the $S(4)$ classification notation defined earlier. The complex internal structure is further evidenced by the cross-sectional spin profile in **Fig 3**(c), where the OP spin component reverses six times along the central transect—a signature of the multiple domain walls stabilized by the substantial DMI. In addition to the $S(4)$, individual skyrmions also appear in the ground state of monolayer CrInTe$_2$. Furthermore, we observe that these skyrmion bags coexist with elementary isolated skyrmions, which exhibit a significantly stronger spatial confinement.

We subsequently explore the influence on the evolution of these topological spin textures under an external magnetic field. As the field increases, the isolated skyrmions undergo a uniform contraction due to the Zeeman energy penalty. The response of the S(4) is far more dramatic: at a relatively low field of 0.1 T, the bag undergoes a topological fission, transforming into two individual skyrmioniums. Structurally, a skyrmionium consists of a target skyrmion nested within a larger skyrmion of opposite polarity. Consequently, it possesses a net topological charge $Q = 0$. This global charge neutrality suppresses the skyrmion Hall effect, making the skyrmionium an ideal candidate for achieving straight-line spin transport. Despite having zero net topological charge, the skyrmionium remains robust stability due to its local topological protection.

Upon further increasing the magnetic field to 0.5 T, the skyrmioniums anaihilate, leaving high-density phase of compact elementary skyrmions. Therefore, the external magnetic field drives a discrete topological hierarchy: from skyrmion bags to skyrmionium and finally to elementary skyrmions. This cascade phase transition demonstrates that the magnetic field can directly and deterministically modulate the topology of the system.

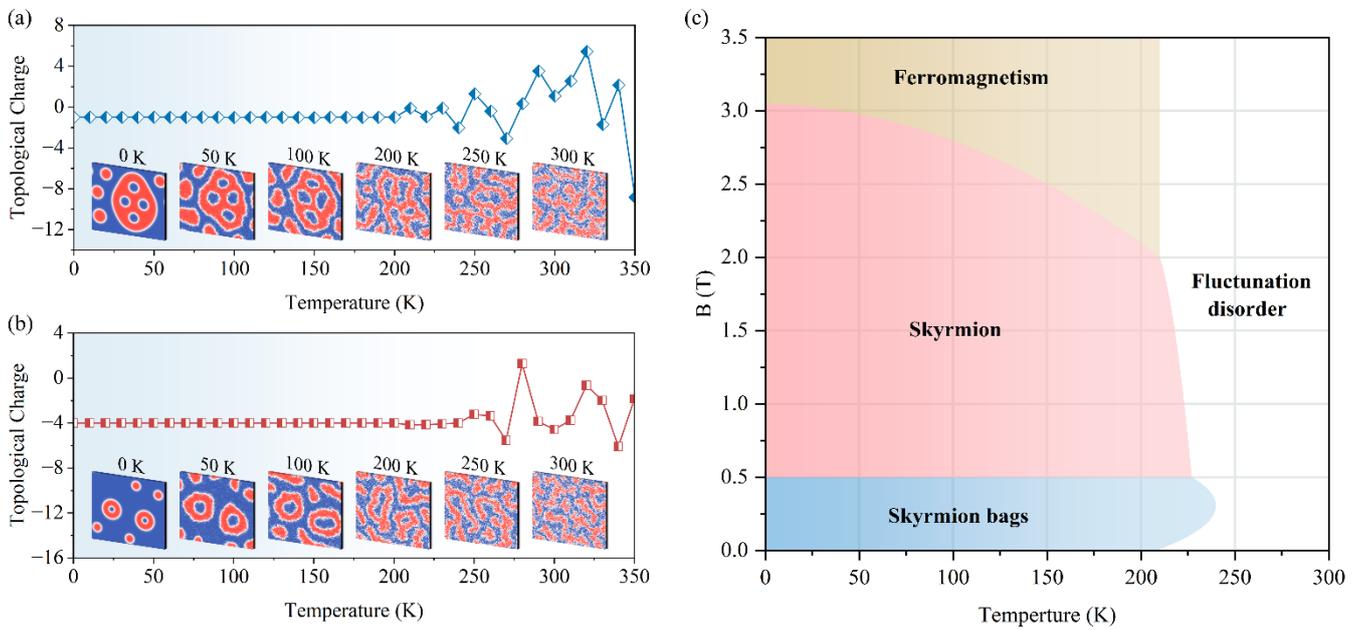

**Fig 4** Evolution of total topological $Q$ and representative spin texture snapshots in monolayer CrInTe$_2$ as a function of temperature under a constant magnetic field of (a) 0 T and (b) 0.1T. (c) Magnetic field-temperature (B-T) phase diagram of monolayer CrInTe$_2$, delineating the stability regions of skyrmion bags and elementary skyrmions.

Given the critical importance of thermal stability for practical applications, we systematically examine the robustness of these topological textures against thermal fluctuations. **Fig 4**(a) and (b) track the evolution of the S(4) state (at B = 0 T) and the skyrmionium (at B = 0.1 T) as a function of temperature. For the zero-field S(4) bag, the topological $Q$ remains strictly quantized, and the composite structure preserves its integrity up to 200 K. As the temperature rises, thermal broadening gradually blurs the domain boundaries, yet the topology remains protected. When the temperature exceeds 200 K, however, significant thermal agitation triggers fluctuations in $Q$, marking a transition into a fluctuation-disordered (FD) state. Remarkably, the field-stabilized skyrmionium (B = 0.1 T) demonstrates even higher thermal robustness, persisting stably up to 240 K. This indicates that the external field provides an additional pinning potential that counteracts thermal randomization. Beyond 240 K, $Q$ starts to fluctuate and the skyrmioniums vanish. These results demonstrate that topological magnetic states in monolayer $CrInTe_2$ can be maintained over a broad temperature range.

To further investigate the stability of topological magnetism in monolayer $CrInTe_2$, we constructed a temperature-magnetic field phase diagram, as shown in **Fig 4**(c). Based on the topological invariants, the phase diagram is demarcated into four distinct regimes: the skyrmion bags phase, the isolated skyrmion phase, the FM state and the FD state. The skyrmion bags phase emerges in a wide range of 0-0.5 T and 0-240 K. Specifically, monolayer $CrInTe_2$ hosts a rich sequence of states in this regime: with increasing external magnetic field, the system evolves sequentially from the $S(4)$ configuration to $S(3)$ and $S(2)$, culminating in the $S(1)$ bag, as shown in **Fig S**7. Beyond this range, isolated skyrmions are observed across a broad region of 0.5–3.05 T and 0–220 K. Notably, in the high-field range of 2.0–3.05 T, the skyrmions become spatially compressed and sparse; consequently, they are more vulnerable to thermal fluctuations, leading to their annihilation into the trivial FM state at relatively lower temperatures.

**Conclusions**

In summary, we establish a general design principle for engineering skyrmion bags with tunable topology in symmetry-broken 2D FM lattices. By elucidating the delicate energetic competition among isotropic exchange, DMI and magnetic anisotropy, we identified a robust stability window for high-order topological states. Applying this framework to monolayer $CrInTe_2$ we validate the existence of spontaneous, field-free skyrmion bags and demonstrate their deterministic evolution into topological neutral skyrmioniums under weak external fields. Notably, the realization of skyrmioniums with remarkable thermal stability up to 240 K offers a practical avenue for achieving efficient spin transport immune to the skyrmion Hall effect. Our findings greatly enrich the research on high-order topological magnetism.

**Supporting Information**

The supporting information is available in ***.

**Acknowledgements**

This work is supported by the National Natural Science Foundation of China (No. 12274261) and Taishan Young Scholar Program of Shandong Province.